\documentclass[twoside]{ilcws08}
\usepackage[latin1]{inputenc}
\usepackage[dvips]{graphicx,epsfig,color}
\usepackage{wrapfig,rotating}
\usepackage{amssymb,amsmath,array}

\newcommand{\beq}{\begin{equation}}
\newcommand{\eeq}{\end{equation}}
\newcommand{\bea}{\begin{eqnarray}}
\newcommand{\eea}{\end{eqnarray}}

\begin{document}
\title{
Top Quark Physics at the ILC: Methods and Meanings} 
\author{Zack Sullivan
\vspace{.3cm}\\
Illinois Institute of Technology - BCPS\\
3101 S.\ Dearborn St., Chicago, IL 60616-3793 - USA
}

\maketitle

\begin{abstract}
The physics case for studying top-quark physics at the International Linear
Collider is well established.  This summary places in context the top-quark
physics goals, examines the current state-of-the art in understanding of the
top-quark mass, and identifies some areas in which the study of the top-quark
mass enhances our understanding of new techniques.
\end{abstract}

\section{Introduction}

The measurement of top-quark production and decay at the International Linear
Collider (ILC) will provide precision access to properties of the top quark,
as well as to multiple signals of physics beyond the Standard Model.
Recently, a strong case has been made for study of top-quark production at the
ILC~\cite{ILCV2}.  Scans at both threshold for $t\bar t$ production, and in
the continuum, provide an opportunity to search for both direct and virtual
effects of physics Beyond the Standard Model (BSM).

Arbitrary couplings of the form $\gamma/Z-t-t$ can be measured to the 1\%
level \cite{Rindani:2003av,Grzadkowski:2002gt}, with improvements beyond
coming from additional positron polarization, or collider energy.  The
left-handed coupling of between $W-t-b$ might be measured to
3\%~\cite{Batra:2006iq}, though beamstrahlung and ISR effects may reduce this
reach~\cite{BoogertTalk}.  These couplings are relevant in models with new
generations, top-flavor, Little Higgs with $T$-parity, and more.  With as
little as 500~$\mathrm{fb}^{-1}$ of data at $\sqrt{S}=500$ GeV, vector-like
top quarks $T$ can be indirectly probed through their modification of the
$Z-t-t$ up to 1~TeV~\cite{Berger:2005ht}.  Many scenarios have been considered
in detail, and are summarized in the Reference Design Report~\cite{ILCV2}.

Methods for obtaining the top-quark Yukawa coupling, width, and mass have been
are also described in detail in the Design Report.  However, the last year has
seen significant improvement in the predictions of event distributions used to
extract the top-quark mass.  As this is where the interest has been, I focus
here (and in my talk~\cite{ZStalk}) on reexamining issues regarding the
top-quark mass, from where we have come, to where we are going.

\section{Top-quark mass: What is it?}

The primary focus of top-quark studies at the ILC has been on developing
methods to extract the most precise top-quark mass
possible~\cite{ILCV2,LHCLC}.  Before addressing the question of \textit{why}
we need a precise top-quark mass, we should be certain we understand
\textit{what} it is we are trying to find, and how we plan to find it.

As the first evidence of the top-quark came to light, it was clear that the
top-quark width was small, and, hence, the top quark decays before it has time
to hadronize.  Given the prospect of probing a quark directly, the question
arose: what is the top-quark mass?  Several reasonable responses are: a
parameter of the Lagrange density (${\cal L}\sim m\bar tt$), an effective
Yukawa coupling between a Higgs boson and two top quarks ($m_t =
Y_t/(2\sqrt{2}G_F)^{1/2}\approx 1$ in the Standard Model), or even the
kinematic mass seen by experiments.  The ``pole mass'' of the top quark was
quickly shown to be ambiguous at the order of $\Lambda_\mathrm{QCD} \sim
150~\mathrm{MeV}$ from a calculation of renormalon
corrections~\cite{Smith:1996xz}.  Unfortunately, the experimental expectation
for extraction of the top-quark mass at the ILC is $100$~MeV.  Hence, there
was a theoretical declaration that the $\overline{\mathrm{MS}}$ mass, which
suffers much smaller ambiguities, should be used as the standard.

The $\overline{\mathrm{MS}}$ mass is useful for reporting results.  However,
masses are not actually measured directly in experiments.  What is measured
are distributions of events from which a mass is \textit{inferred}.  The two
distributions at the ILC currently under review are the 1S (``threshold'')
mass, and the top-quark jet mass (``continuum'').  Distributions are fit using
these proxy-masses; and perturbative expansions are used to work back to the
$\overline{\mathrm{MS}}$ mass.  Recent progress in predicting the
distributions to be fit in the data was presented at this
conference~\cite{BoogertTalk,SeidelTalk,MantryTalk}.

The primary method of extracting the top-quark mass at the ILC is to fit a
theoretical shape prediction to a scan across the $t\bar t$ threshold.  An
early study pointed out the theoretical prediction of the shape was unstable
when using the pole mass (or $\overline{\mathrm{MS}}$ mass) due to
non-relativistic corrections near threshold \cite{Yakovlev:2000pv}.  Use of
the pseudo bound-state 1S mass allowed a systematic improvement of the
threshold region as a non-relativistic expansion of the cross section of the
form
\beq
\sigma_{t\bar t}\propto v\sum\left(\frac{\alpha_s}{v}\right)\times \left\{\!\!
\begin{tabular}{c} 1\\ {$\sum(\alpha_s\ln v)$}\end{tabular}\!\!
\right\} \times \left\{\!\!\! \begin{tabular}{l} {$\mathrm{LO}(1) +
\mathrm{NLO}(\alpha_s,v) + \mathrm{NNLO}(\alpha_s^2,\alpha_s
v,v^2)$}\\ {$\mathrm{LL} + \mathrm{NLL} +
\mathrm{NNLL}$}\end{tabular}\!\!\! \right\} \;.\label{zs:eq_exp}
\eeq
Normalizations change at each order, but use of the 1S mass provides a fairly
stable threshold peak distribution shape.  This systematic approach leads to
an estimated uncertainty $\delta\sigma_{t\bar t} = \pm 6\%$ before
initial-state radiation (ISR) or beamstrahlung are included --- we'll return
to this.  Based on this alone, it appears extraction of $\delta m_t\sim
100$~MeV is attainable \cite{ILCV2}.

\begin{wrapfigure}{r}{0.5\columnwidth}
\centerline{\includegraphics[width=0.5\columnwidth,clip]{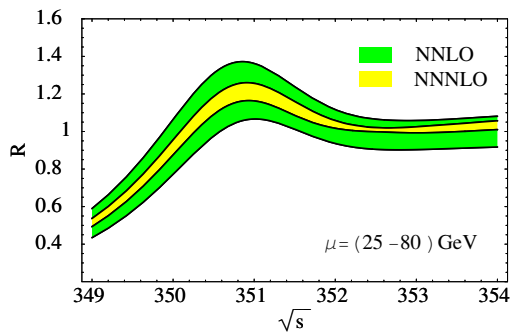}}
\caption{Ratio of NNNLO (and NNLO) to LO corrections across the $t\bar t$
threshold~\protect\cite{Beneke:2008cr}}\label{zs:fig_sigtt}
\end{wrapfigure}

Recent theoretical work has extended these calculations to include even
smaller effects.  Corrections due to the unstable nature of the top quark
\cite{Fleming:2007qr,Hoang:2006pd} have been found to be 3--10\% across the
threshold region.  Figure~\ref{zs:fig_sigtt} shows that most of the NNNLO term
of the expansion in Eq.\ \ref{zs:eq_exp} has now been
calculated~\cite{Beneke:2008cr}.  At this conference we heard about the
electroweak-strong correction of order $\alpha\alpha_s$
\cite{SeidelTalk,Kiyo:2008mh} which is at the level of $\delta\sigma_{t\bar
t}\sim 0.1\%$.  The overall corrections examined here are well below the
expected experimental precision.  Does this mean that everything is ready for
the ILC to turn on?

\begin{wrapfigure}{l}{0.5\columnwidth}
\centerline{\includegraphics[width=0.5\columnwidth,clip]{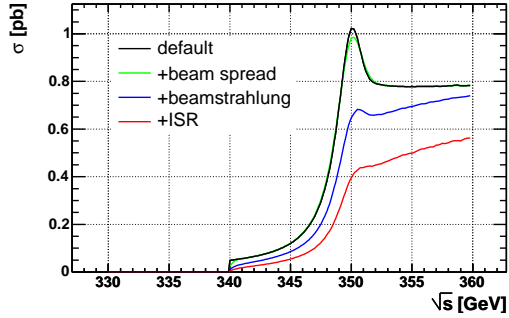}}
\caption{Smearing of the $t\bar t$ threshold from beamstrahlung and ISR
effects~\protect\cite{BoogertTalk}}\label{zs:fig_isr}
\end{wrapfigure}

There remain several effects on the order of 50~MeV each that remain to be
understood.  The most pressing issue will be the inclusion of beamstrahlung
and ISR.  It is apparent from Fig.~\ref{zs:fig_isr}, shown at this
conference~\cite{BoogertTalk}, that beam and ISR effects will dramatically
smooth the threshold region.  It has been suggested \cite{ILCV2} that a
forward tracker that can observe Bhabha scattered events could help model some
of these effects and maintain a $\delta m_t\sim 100$--200~MeV.  What has not
been studied, is how to correctly match these ISR effects to the resummed
calculations such that there is no double-counting of ISR.  I suggest that the
likely the matching uncertainty is at least 50 MeV, and could ultimately
dominate the uncertainty in the top-quark mass extraction.  Significantly more
work is needed to understand how to combine these theoretical cross sections
and ISR at the ILC.

A new approach to extracting a competitive value for the top-quark mass away
from threshold has has grown out of the recent successes of effective field
theories (EFTs) of QCD.  The ``continuum'' top-quark mass is extracted by a
fit to the mass of the jet formed by decay products of the top quark.
Demonstration of a formal factorization of the production and decay process
into strongly ordered scales ($Q \gg m_t \gg \Gamma_t >
\Lambda_{\mathrm{QCD}}$)~\cite{Fleming:2007xt,Hoang:2008fs}, allows for a
thrust-like object called the ``hemisphere'' mass to be constructed.  For
details, see the work of Ref.~\cite{MantryTalk} at this conference.  This
represents the beginning of a new class of calculations that describe a
complete event --- from production through decay and hadronization.

\section{Top-quark mass: Why do we care?}

Now that we have established what it is that we measure, it is useful to
examine in a bit more depth why we care.  Explicitly, why does it matter that
we measure the top-quark mass to 100 MeV, as opposed to 1 GeV?  The typical
reason given for studying the top-quark mass is that it provides a strong
constraint on electroweak physics.  Specifically, since the top-quark and
Higgs boson contribute at one loop to the $W$ and $Z$ propagators, the $W$
mass is quadratically sensitive to the top-quark mass, and logarithmically
dependent on the Higgs mass.  Inverting this dependence leads to constraints
on the Higgs mass based on fits to the electroweak precision data, and the
direct measurement of the $W$ and top-quark masses at the Fermilab Tevatron.
By summer 2008, the top quark was measured to be $172.4\pm
1.2$~GeV~\cite{TevEWWG}.

A better way to look at the Higgs constraint was pointed out in
Ref.~\cite{Beneke:2000hk}.  Assuming that the Higgs mass is known, and $m_W$
is measured to $20$ MeV (the LHC target), our current understanding of
$\sin^2\theta_W$ implies we only need to know $m_t$ to $\sim 3$~GeV at the
LHC.  The ILC can measure $m_W$ to $\sim 6$~MeV~\cite{ILCV2}.  Giga-Z proposes
to measure $\sin^2\theta_W$ to $\sim 10^{-5}$ (see the talk of Ref.\
\cite{MPTalk} for a method to use $Z$-calibration data to attain $3\times
10^{-5}$).  This optimal scenario cannot make use of a top-quark mass
uncertainty better than 1~GeV in the Standard Model, a value already saturated
by the current Tevatron measurement!

The drive to extract an extremely accurate top-quark mass at the ILC, then, is
from physics Beyond the Standard Model.  Many models of new physics are very
sensitive to the exact top-quark mass.  For example, in supersymmetry, the
shift in the Higgs mass is
\beq
\Delta m_H^2\approx \frac{3G_F m_t^4}{\sqrt{2}\pi^2\sin^2\beta}
\ln\left(\frac{\overline{m}_{\tilde{t}}^2}{m_t^2}\right) \;,
\eeq
where $\tan\beta$ is the ratio of vacuum expectation values of the two Higgs
doublets, and $\overline{m}_{\tilde{t}}$ is the average top-squark mass.  In
this case, the theoretical error in the Higgs mass will be directly
proportional to the error in the top-quark mass.  Hence, the desire to reach
$\delta m_t\sim 100$~MeV.  In addition to the Higgs mass, several other
parameters of supersymmetry may be strongly constrained, such as $M_A, A_t,
m_{1/2}$, etc.

Given that we expect some new physics must exist that explains electroweak
symmetry breaking, the motivation to measure a small top-quark mass is strong.
It also provides significant theoretical motivation to do higher-order loop
corrections.  Simple power counting suggests that four-loop corrections will
contribute at the same order numerically as the measurement.  Hence, the
industry of loop calculations in the Standard Model and supersymmetry has a
significant effort ahead of it before we can make use of the phenomenal
precision that will be achieved at the ILC.

\section{Status and remaining questions}

The physics case for an extended study of top-quark production and decay at
threshold and in the continuum is already strong \cite{ILCV2}.  The primary
focus of the last several years has been to improve techniques for extracting
the top-quark mass.  Questions have been raised as to whether the
best-measured mass will be the 1S mass from threshold, or a top-quark jet mass
from the continuum.  In the end, neither mass will be measured.  Rather, we
will measure line shapes or particle flow, or invariant masses with explicit
cuts --- all under the influence of ISR/FSR radiation effects.  Our ability to
extract the top-quark mass under these conditions will, therefore, be
constrained by our ability to understand and control the systematic
uncertainties introduced by our modeling of these phenomena.

This leads to the following questions:
\begin{enumerate}
\item Do we need to work out initial-state subtraction terms to merge formal
calculations correctly with ISR estimates for incoming elections?  Box
diagrams and $\alpha^2$ diagrams may be significant on the level of the
$0.01\%$ theory error we want for $m_t$ and other observables.

\item The hemisphere-like top-jet definitions are exciting theoretical
constructs, but does the factorization demonstrated hold in the presence of
hard cuts?  Once regions are eliminated due to finite-size detector effects,
the current definitions look similar to very large fixed-cone definitions,
which are not infrared-safe.  More work needs to be done to understand whether
experimental implementation of these theoretical constructs is viable.

\item One jet algorithm expected to be used is particle flow.  This procedure
estimates the energy in the neutral particles in the event from the charged
ones.  Does this invalidate any assumptions in the theoretical calculations?
If we answer in the negative, we must be certain to better than $0.01\%$.
\end{enumerate}

The greatest opportunity and challenge going forward will be to ensure that
the experimental and theoretical definitions agree.  Here, a significant
opportunity presents itself in the application of the new effective field
theories.  The merging of perturbation theory (PT), soft-collinear effective
theory (SCET), heavy-quark effective theory (HQET), and other derivative
theories that better describe jet structure, will produce the tools we need to
achieve a theoretical precision in QCD comparable to the amazing precision
that will be obtained experimentally at the International Linear Collider.



\begin{footnotesize}

\end{footnotesize}


\end{document}